# Causal Inference on Investment Constraints and Non-stationarity in Dynamic Portfolio Optimization through Reinforcement Learning

Yasuhiro Nakayama[1], Tomochika Sawaki [2]

[1] Mizuho Research & Technologies, Ltd.
[2] Mizuho Bank, Ltd.
yasuhiro.nakayama@mizuho-rt.co.jp, tomochika.sawaki@mizuho-bk.co.jp

**Abstract:** In this study, we have developed a dynamic asset allocation investment strategy using reinforcement learning techniques. To begin with, we have addressed the crucial issue of incorporating non-stationarity of financial time series data into reinforcement learning algorithms, which is a significant implementation in the application of reinforcement learning in investment strategies. Our findings highlight the significance of introducing certain variables such as regime change in the environment setting to enhance the prediction accuracy. Furthermore, the application of reinforcement learning in investment strategies provides a remarkable advantage of setting the optimization problem flexibly. This enables the integration of practical constraints faced by investors into the algorithm, resulting in efficient optimization. Our study has categorized the investment strategy formulation conditions into three main categories, including performance measurement indicators, portfolio management rules, and other constraints. We have evaluated the impact of incorporating these conditions into the environment and rewards in a reinforcement learning framework and examined how they influence investment behavior.

## 1. Introduction

In recent years, remarkable progress has been made in research and development applying machine learning to investment strategies. These research directions can be broadly categorized into two types. The first category involves reading or interpreting unstructured data, such as text and images, using models built by deep learning. The second category encompasses using machine learning models to estimate parameters that were previously estimated by linear or statistical models in financial engineering.

The advantage of the latter model of machine learning is that it can be extended to non-linear and more complex models, such as deep learning and ensemble machine learning, thus suggesting the possibility of estimating parameters in higher dimensions than conventional models. This leads to the construction of models with higher operational performance. Typically, the investment decision-making process of a portfolio manager, using the quant method, can be segmented into three steps: data collection and processing, analysis and signal generation, and portfolio optimization. The first step is related to data collection and processing, and the second step is related to analysis and signal generation. In this study, our focus is on the potential application of machine learning and artificial intelligence technologies in the formulation stage of investment decision making, which corresponds to the third step.

In particular, portfolio management by institutional investors, also known as asset owners, is a part of the larger goal of enhancing enterprise value. In such circumstances, it becomes necessary to search for an optimal portfolio for the company based on a multifaceted understanding of financial accounting, legal and regulatory affairs, and taxation, rather than aiming for a theoretically efficient portfolio. Therefore, this study examines the utility of reinforcement learning methods for formulating optimal portfolios incorporating such practical constraints.

## 2. Related research

Reinforcement learning is a field of machine learning that is designed to learn sequential decision rules. The unique feature of reinforcement learning is that it rewards objectives and learns how to achieve those objectives from the data without presuming complete knowledge about the system or environment to which it is applied [1].

In investment decision making, the primary consideration when using reinforcement learning pertains to the modeling of non-stationarity in financial time series data. This implies that mean, variance, and covariance are not constant over time and that several events have been historically observed, such as sudden shifts in phase or spikes in volatility. A specific approach to addressing non-stationarity in reinforcement learning has been described in [2]. Furthermore, research studies focusing on non-stationarity in financial time series data and employing reinforcement learning in asset allocation are included in [3-16].

3 Method
3.1 Data Preprocessing

In this study, we examine dynamic optimal allocation through decision making for rebalancing two assets, namely, a risky asset and a non-risky asset, using daily data from April 2000 to March 2023. We choose the total return index of the S & P500, denominated in USD, as the risky asset and the total return index of US Treasury Index, also denominated in USD, as the non-risky asset. The data is sourced from Bloomberg, and the daily return is calculated for the analysis. We don't take into account funding costs, transaction costs, and cash ratios.

3.2 Learning Techniques
(1) learning technique

In common with Chapters 4 and 5, the process of reinforcement learning model learning in this study is as follows. Reinforcement learning involves maximizing the sum of rewards through the observation of states from the environment and the corresponding action. To analyze the behavior of each state, we utilize table-based reinforcement learning methods such as SARSA and Q-learning.

<SARSA Update Expression >
$$Q'_\pi(S_t, A_t) = Q_\pi(S_t, A_t) + \alpha\{R_t + \gamma Q_\pi(S_{t+1}, A_{t+1}) - Q_\pi(S_t, A_t)\}$$

<Q Learning Update Expression >
$$Q'(S_t, A_t) = Q(S_t, A_t) + \alpha\left\{R_t + \gamma \max_a Q(S_{t+1}, a) - Q(S_t, A_t)\right\}$$

where $Q$ is the action value, $R$ is the immediate reward, $S$ is the state, $A$ is the action, $\alpha$ is the learning rate, and $\gamma$ is the discount rate. The ε-greedy method is used to study 1,000 episodes for each study period.

(2) behavior

Each year, starting with 50% risky assets and 50% non-risky assets, the allocation is changed by 10% on a daily basis. There are three possible actions: [risky assets: +10%, non-risky assets: -10%] [no weight change] [risky assets: -10%, non-risky assets: +10%]. However, if the weight has already reached 100%, the weight will not change even if it chooses to increase the weight.

(3) remuneration

The basic compensation is determined by taking the difference between the Sharpe ratio of the portfolio and the Sharpe ratio of a benchmark portfolio for both the year-to-date and the last 10 days, and then summing these differences to obtain the basic compensation. For the benchmark portfolio, the Sharpe ratio is determined at the highest allocation with fixed weights for both risky and non-risky assets for each year. This kind of allocation is regarded as a correct allocation that is ascertainable retrospectively.

The Sharpe ratio is calculated as follows: $SR$

$$SR = \frac{r}{\sigma}$$

where $r$ is the portfolio return and $\sigma$ is the standard deviation of the portfolio return. The risk-free rate is 0.

The basic reward at time t is given by the following formula: $R_t^{Base}$

$$R_t^{Base} = (SR_t - SR_t^{BM}) + (SR_t^{10} - SR_t^{BM,10})$$

$SR_t^{BM}$ is the Sharpe ratio of the benchmark portfolio, and $SR_t^{10}$ is the Sharpe ratio calculated from the returns over the last 10 days.

In each analysis, a different reward may be added to the basic reward, which will be described in detail later.

(4) State

The status of each analysis case is different and will be described later.

4. Out of sample back-test considering regime change
4.1 Analytical Methods

In this chapter, we investigate how performance prediction using reinforcement learning models is influenced by the non-stationarity of financial time series data, particularly when the mean and variance change over time. We consider two models. The first model solely defines the state space in terms of the expected returns of risky and non-risky assets. For the expected return, we simply use the price difference (momentum) from 60 business days ago. Binary classification based on the positive and negative momentum of the asset is performed, and the state space is divided into four states of 2x2. As for the second model, it adds the correlation coefficient between the risky asset and the non-risky asset to the state variable. The correlation coefficient is estimated from daily data for the past 60 business days. The correlation coefficient state variable is divided into three states of positive correlation/no correlation/negative correlation with a threshold of ±0.2 and is combined with the expected return, resulting in a total of 12 states. SARSA is used in this chapter. The first model is referred to as the base model, the second model as the non-stationary model and the back-test's evaluation is also compared to a random model using a Q-table as a random variable. The random model generates random Q-tables 1000 times to measure performance.

4.2 Back-test Results

During the learning period, the entire fiscal year from April to March of each year is used as a set to estimate the Q-table for each set. For out-of-sample back-test validation, Q-tables estimated in the past learning period rather than the validation period are utilized, and those in the future period are not applied. For example, when conducting out-of-sample back-test verification from April 2018 to March 2019, the average value of each element of the total 18 Q-tables estimated for the study period from FY 2000 to FY 2017 is calculated using equal weights. Additionally, the rebalancing frequency is daily, and transaction costs are not considered.

Table 1 compares the out-of-sample performance for each year with the median of the base model, the non-stationary model, and the random model. Back-test performance is defined by the Sharpe ratio, which is the annual return divided by the annual standard deviation. The non-stationary model's average Sharpe ratio from FY2001 to FY2022 is higher than that of the base model, and the difference is statistically significant. The non-stationary model also outperforms the median of the random model on average Sharpe ratios. Comparing the results of each year, the first half of the back-test period, i.e., the 2000s, have more years in which the non-stationary model performs worse than the base model than the second half of the back-test period, i.e., the 2010s. This could be due to the learning period beginning in FY 2000, whereby sufficient time to forecast the future is not provided.

Figure 1 compares the probability density of the Sharpe ratio of the random model with that of the random Q-table. The purpose of this analysis is to assume that there are some years in which the performance of an asset returns significantly depending on the shape of its short-term momentum and daily change rate distribution within the year, and some years in which it does not. For example, an event (such as a presidential election or during a significant event, such as a FOMC meeting), the probability density of a random model is known to be bimodal, and annual performance changes dramatically depending on whether the portfolio holds a significant weight of risky assets at that time.

Table 1: Sharpe ratio for each year of back-testing

| FY | Base Model | Random Model (Median) | Non-Stationary Model |
|---|---|---|---|
| 2001 | 0.42 | 0.16 | 0.59 |
| 2002 | 2.21 | -0.69 | 1.38 |
| 2003 | 0.97 | 1.75 | 0.36 |
| 2004 | 0.33 | 0.48 | -0.01 |
| 2005 | 0.62 | 0.95 | 1.28 |
| 2006 | 2.05 | 1.15 | 2.03 |
| 2007 | 2.34 | 0.19 | 2.74 |
| 2008 | 1.08 | -0.01 | 1.08 |
| 2009 | 0.01 | 1.63 | 0.01 |
| 2010 | 1.01 | 0.77 | 1.16 |
| 2011 | 1.72 | 0.79 | 1.72 |
| 2012 | 0.85 | 1.01 | 0.85 |
| 2013 | -0.49 | 1.25 | 1.59 |
| 2014 | 1.61 | 1.30 | 1.64 |
| 2015 | 0.57 | 0.03 | 1.85 |
| 2016 | -0.49 | 1.10 | -0.37 |
| 2017 | 0.06 | 0.79 | 1.88 |
| 2018 | 1.62 | 0.81 | 0.81 |
| 2019 | 2.04 | 0.16 | 2.00 |
| 2020 | -0.74 | 1.60 | 1.65 |
| 2021 | -0.73 | 0.53 | -0.12 |
| 2022 | -0.56 | -0.45 | 0.61 |
| Average(All) | 0.75 | 0.70 | 1.12 |
| Average(2005〜) | 0.70 | 0.76 | 1.24 |

Figure 1: Results for the last 10 years (horizontal axis: Sharpe ratio)

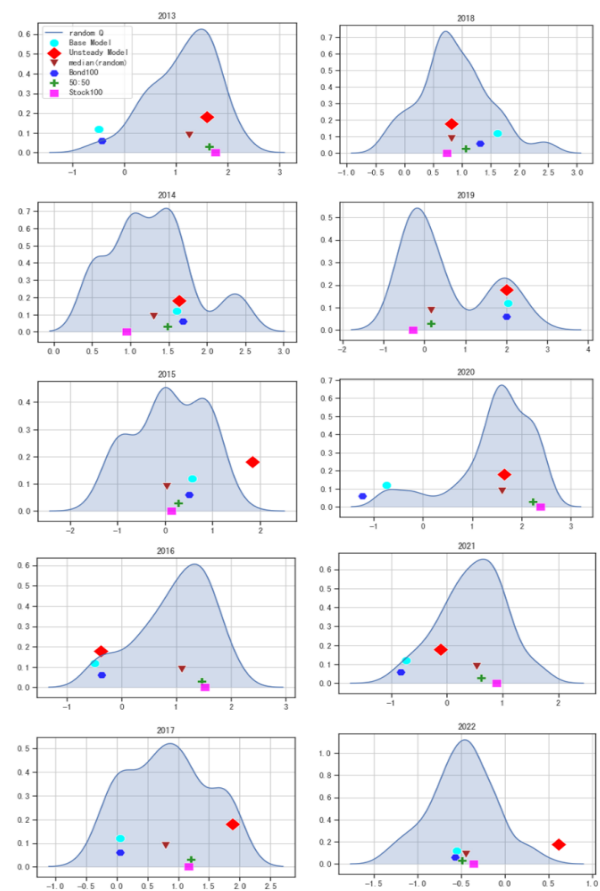

*Bond100 = 100% non-risky assets, 50:50 = 50% risky assets: 50% non-risky assets, Stock100 = 100% risky assets Buy & hold for 1 year

In such a case, the non-stationarity of financial time series data strongly emerges, confirming the effectiveness of the non-stationary model. As shown in Table 1, it was observed that the Sharpe ratio of the non-stationary model exceeded the median Sharpe ratio of the random model in more than half of the years. Moreover, it was confirmed that in years with a bimodal shape, the non-stationary model generally had fewer years within the range of the lower peak of the Sharpe ratio. These results suggest that adding a phase to the state variable of a reinforcement learning model contributes to a significant improvement in prediction accuracy.

## 5. Comparison of decision making considering investment constraints

### 5.1 Analytical Methods

In this chapter, we examine how the behavior selected in reinforcement learning transforms and impacts performance when considering practical restrictions set in chapter 3 while the previous chapter focuses on prediction effectiveness, this chapter compares the results of in-sample learning to analyze the relationship between different constraints and decision making. Table 4 illustrates three categories of indicators and rules to consider when optimizing portfolios at each time point. The first category is a kind of performance measurable indicators. For instance, besides the target returns and Sharpe ratios for the fiscal year, risk indicators such as VaR (value-at-risk) and drawdown, and management indicators related to portfolio risk returns such as loss cut points are utilized. The second category comprises rules and regulations associated with portfolio management. For instance, rules related to periods such as investment horizons and rebalancing frequency, regulations related to settlement such as margin and clearing, regulations related to financial indicators such as leverage ratio, risk weighted assets, and liquidity ratio, and regulations related to various financial supervisory authorities, such as the Volcker rule, are assumed. The third category comprises the constraints that can be considered, such as the accuracy of expected returns and transaction costs.

Table 4: Examples of Indicators and Rules Considered in Portfolio Optimization

| Category | Example |
| --- | --- |
| ① Performance Management Indicators | Target return, target volatility, risk indicators (VaR, drawdown, etc.), loss cut point, etc. |
| ② Portfolio Management Rules | Period, Settlement, Finance, Regulation, etc. |
| ③ Other Constraints | Information sources, expected return accuracy, transaction costs, etc. |

In this chapter, we establish a benchmark portfolio and compare its performance with that of learning with each constraint. Before conducting back-testing, we evaluate the changes in performance and behavior resulting from the signal accuracy concerning category (3) shown in Table 4. A binary signal is generated using the answer to whether the Sharpe ratio of the risky asset or the non-risky asset is higher for the next five business days, and the status is set accordingly. Additionally, we explore the case where a certain percentage of the signal is reversed (i.e., substituted with the wrong signal). The rate of inversion is classified into steps of 10%, and six patterns are examined until the true positive: false =50:50 to ascertain the performance change. Figure 2 depicts signal accuracy and performance. As in earlier sections, the years ranging from FY 2000 to FY 2022 are analyzed one by one, and the distribution of Sharpe ratios obtained each year is plotted in the in-sample. In addition to the monotonic increase in the median Sharpe ratio as the horizontal axis moves right (as signal accuracy increases), the distribution width expands as the signal's accuracy approaches 100%. This implies that the Sharpe ratio expected to be obtained may nonlinearly increase as the information's dominance and signal accuracy increase.

Figure 2: Signal Accuracy and Performance

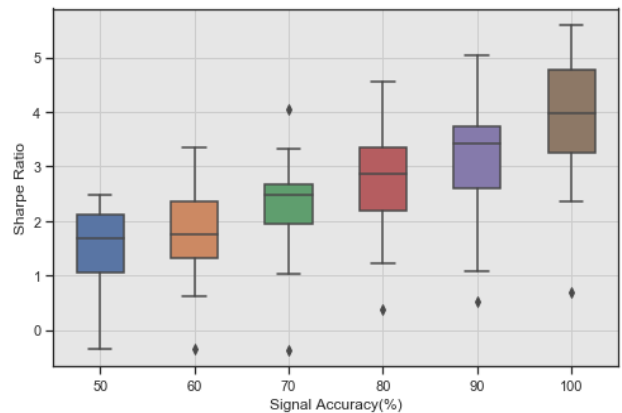

### 5.2 Back-test Results

First, a total of seven kinds of back-tests were set for category (1) shown in Table 4, as shown in Table 5 below, and changes in behavior are compared. In this chapter, we use Q-learning.

Table 5: List of Back-test Settings for Category (1) Performance Management Metrics

| No. | state variables | rewards |
|---|---|---|
| #001 | signal, position, age | basic reward |
| #002 | Signal, Position, Quarterly, Target Achievement Status (1 step) | basic reward + Target Achievement Reward (1 level) |
| #003 | Signals, Position, Quarterly, Target Achievement Status (2 levels) | basic reward + Target Achievement Reward (2 levels) |
| #004 | Signal, Position, Quarterly, DD Occurrence (1 level) | basic reward +DD penalty (one level) |
| #005 | Signal, Position, Quarterly, DD Occurrence (2 levels) | basic reward +DD penalty (2 levels) |
| #006 | Signal, Position, Quarterly, Target Achievement Status (1 step), DD Occurrence Status (1 step) | basic reward + Target achievement reward (one level), +DD penalty (one level) |
| #007 | Signal, Position, Quarterly, Target Achievement Status (2 Stages), DD Occurrence Status (2 Stages) | basic reward + Target Achievement Reward (2 levels) +DD penalty (2 levels) |

We utilize a binary signal to determine whether the observed Sharpe ratio of the risky asset or the non-risky asset performs better. With an Accuracy of 60%, we use the Accuracy 60% signal to examine the behavior learned by reinforcement learning in a probabilistic situation where signal effectiveness varies. Depending on the current holding ratio, three positions are generated: high ratio of risky assets/equal ratio of non-risky assets/high ratio of non-risky assets. This is because the meaning of the "no weight change" behavior depends on the current overweight asset. We divided the year into four equal parts or quarters as the elapsed period and created four conditions.

We analyze the selected actions resulting from learning. [No weight change] is chosen as a preference for risky assets when the risky asset ratio is elevated and as a preference for non-risky assets when the non-risky asset ratio is elevated. (The calculation is excluded when No weight change is selected in the equal-weight state.) We combined the results of all 23 years to establish which behaviors are chosen in each condition and the corresponding rates.

Our base case is defined as back-test number #001, and the difference from the base case is primarily examined. Figure 3 illustrates the percentage of risky assets selected when the signal shows better performance of risky assets and deducts the percentage of risky assets selected when the signal indicates better performance of non-risky assets.

Figure 3: Signals and Behavior

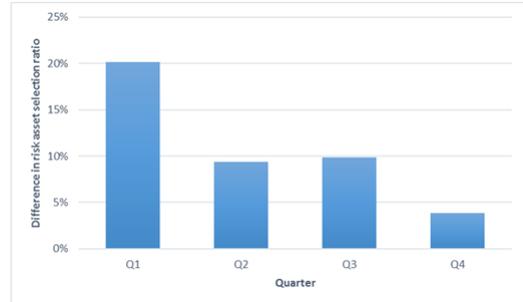

In all quarters, when the signal indicated the performance of the risky asset, the reinforcement learning algorithm is more inclined to accumulate the risky asset, demonstrating that it progresses as per the signals. Additionally, it is noted that the difference tends to decrease as the period progressed.

Next, we examine the ratio of the result of #002 by comparing it to the base case. The target achievement status in the state variable is defined as a binary value of whether the cumulative return from the beginning of the period exceeded 5%. Likewise, an additional reward (+1) is added if the cumulative return exceeds 5% from the beginning of the period. Analyzing the difference in the rate of action from the base case reveals that, in Q3 and Q4, the rate of accumulation of risky assets decreases when the target is achieved, and the signal don't suggest the performance of risky assets. If they remain in this state, they would receive additional rewards. Therefore, they might have learned that it is preferable to avoid risks by maintaining this state rather than taking additional risks by returning to a state where the target has not been attained. This outcome aligns with the expectation that it is better to take a risk-averse approach towards the final half of the year when the target is achieved.

Moreover, it is observed that when targets are not achieved, the proportion of risky assets selected increased, particularly in the fourth quarter. This implies that to attain an additional reward for achieving the target, they learn to favor risky assets with high volatility, that is, large price ranges, aiming to achieve the target within a limited remaining time.

Table 6. Change in Risky asset Preferences #002

| | Signal: Risky assets outperform | | Signal: Non-risky assets outperform | |
|---|---|---|---|---|
| | otherwise | target(5%) achieved | otherwise | target(5%) achieved |
| Q1 | 3% | 6% | 17% | 23% |
| Q2 | 15% | 8% | 8% | 9% |
| Q3 | 3% | -3% | 12% | -4% |
| Q4 | 14% | 10% | 21% | -1% |

In #003, we examine the situation where a goal is defined in two steps. For state variables and rewards, we set a two-stage target of 5% and 10% cumulative returns from the beginning of the period, with additional compensation of +1 at 5% and +2 at 10%. Analyzing the difference in behavior from the base case reveals that the proportion of selecting risky assets increased, even when the first-stage target is achieved, compared to #002. It is suggested that by providing incentives in the form of additional rewards even after achieving the goal, it is possible to make decisions without reducing the risk preference.

Table 7. Change in Risky asset Preferences #003

| | Signal: Risky assets outperform | | | Signal: Non-risky assets outperform | | |
|---|---|---|---|---|---|---|
| | otherwise | target(5%) achieved | target(10%) achieved | otherwise | target(5%) achieved | target(10%) achieved |
| Q1 | 5% | -1% | -5% | 19% | 22% | 20% |
| Q2 | 8% | 6% | 5% | 11% | 11% | 31% |
| Q3 | 15% | 0% | 0% | 9% | 4% | -2% |
| Q4 | 13% | 19% | 14% | 22% | 25% | 9% |

Next, we discuss drawdown in #004. Similar to the achievement of target #002, the cumulative return from the beginning of the period is measured, and two values are added to the state variable, indicating whether the drawdown exceeds 5%. Whenever it exceeds, the reward is penalized (-1). Comparing the rate of action from the base case revealed that the proportion of choosing to add risky assets declined when the drawdown don't happen, even when it indicates the performance of risky assets. This phenomenon is prominent in the early to middle parts of the back-test period. This change could be attributed to a shift towards risk-averse decision-making to avoid the drawdown penalty.

Table 8. Change in Risky asset Preferences #004

| | Signal: Risky assets outperform | | Signal: Non-risky assets outperform | |
|---|---|---|---|---|
| | otherwise | drawdown exceeded 5% | otherwise | drawdown exceeded 5% |
| Q1 | -15% | -1% | 16% | 28% |
| Q2 | -3% | 15% | 16% | 4% |
| Q3 | -18% | -18% | -4% | -4% |
| Q4 | 6% | 5% | 7% | 2% |

In #005, we analyze the scenario where drawdown is set to two levels. The cumulative return of the back-test from the beginning of the period is evaluated, and three values are added to the state variable, indicating whether the drawdown exceeds 5% and 10%. Additionally, the penalty adds to the reward is also set to two levels. The extra penalties are -1 for 5% and -2 for 10%. Compared to the base case, the proportion of selecting risky assets increases beyond the second drawdown, regardless of the signal. This implies that when performance deteriorates significantly, decisions are inclined towards favoring risky assets with high volatility to avoid penalties.

Table 9. Change in Risky asset Preferences #005

| | Signal: Risky assets outperform | | | Signal: Non-risky assets outperform | | |
|---|---|---|---|---|---|---|
| | otherwise | drawdown exceeded 5% | drawdown exceeded 10% | otherwise | drawdown exceeded 5% | drawdown exceeded 10% |
| Q1 | -7% | -1% | 2% | 7% | 19% | 16% |
| Q2 | -5% | -8% | 9% | 0% | 10% | 16% |
| Q3 | -15% | -11% | -6% | -12% | -9% | 0% |
| Q4 | 3% | 6% | 11% | 8% | 2% | 16% |

In #006, both the target achievement status and the occurrence status of drawdown are considered. Analyzing the conditions in which the signal indicates the performance of the risky asset, the target has not been achieved, and no drawdown has occurred reveals that the proportion of selecting risky assets is lower than the base case in the first half of the back-test period, while it increases in the second half. This indicates that avoidance of penalties is prioritized during a certain period, resulting in a preference for risk-averse decisions from the beginning of the period. However, towards the latter half of the period, priority is given to achieving the goals, leading to a transition to risk-preference decision-making.

Table 10. Change in risky asset preference ratio #006

| | Signal: Risky assets outperform | | | Signal: Non-risky assets outperform | | |
|---|---|---|---|---|---|---|
| | target(5%) achieved | otherwise | drawdown exceeded 5% | target(5%) achieved | otherwise | drawdown exceeded 5% |
| Q1 | -8% | -3% | 2% | 14% | 14% | 19% |
| Q2 | -2% | -6% | 9% | 20% | 4% | 10% |
| Q3 | 1% | 7% | 6% | -7% | 16% | 9% |
| Q4 | 10% | 30% | 21% | -5% | 39% | 20% |

Finally, in #007, we set two levels for both the target achievement status and the drawdown occurrence status. The results of this action selection demonstrate the same behavior as when the goal and the drawdown are separately set, either in two stages or in single stages.

Table 11. Change in risky asset preference ratio #007

| | Signal: Risky assets outperform | | | | | Signal: Non-risky assets outperform | | | | |
|---|---|---|---|---|---|---|---|---|---|---|
| | target(10%) achieved | target(5%) achieved | otherwise | drawdown exceeded 5% | drawdown exceeded 10% | target(10%) achieved | target(5%) achieved | otherwise | drawdown exceeded 5% | drawdown exceeded 10% |
| Q1 | 4% | -8% | -2% | 2% | 6% | 19% | 19% | 10% | 15% | 23% |
| Q2 | 10% | -4% | 4% | 8% | 9% | 9% | 7% | 13% | 15% | 11% |
| Q3 | -11% | -14% | 7% | 6% | -15% | 3% | -6% | 15% | 0% | -13% |
| Q4 | 3% | 15% | 14% | 15% | 8% | -4% | 2% | 25% | 11% | 13% |

Next, we compare the categories (2) illustrated in Table 4, focusing on the rebalancing frequency constraint as the simplest constraint. We compare four back-test scenarios: daily, weekly, biweekly, and monthly rebalancing, with no restrictions on rebalancing frequency. The results are presented in Figure 4. As expected, the unconstrained condition exhibits the highest average Sharpe ratio, and performance gradually decreases as the interval is extended.

Figure 4: Rebalancing constraints

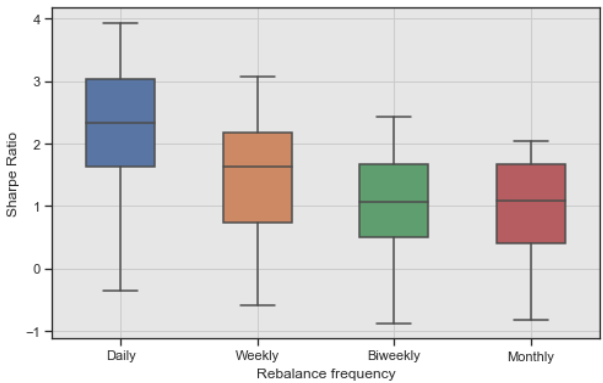

We subsequently analyze the impact of signal accuracy on decision-making in relation to category (3). Similar to #002, we vary the signal Accuracy and compare the selected behaviors. Figure 5 illustrates the changes in signal accuracy and the proportion of assets selected in response to the signals. Each is compared to a signal with an Accuracy 50% (i.e., random). As targets are seldom met in the initial half of the fiscal year, calculations are limited to 3Q and 4Q only. The rate of selection for non-risky assets increases as signal accuracy improved, whereas for risky assets, the trend is discerned in the order of each Accuracy, albeit it is weaker than that for non-risky assets. Based on these findings, it can be assumed that the model is able to objectively assess the accuracy of the given signal and make appropriate decisions. The weaker trend for risky assets could be attributed to the fact that the reward for additional risk-taking is deemed to be smaller than the reward for target achievement.

Figure 5: Changes in signal accuracy and asset selection ratio

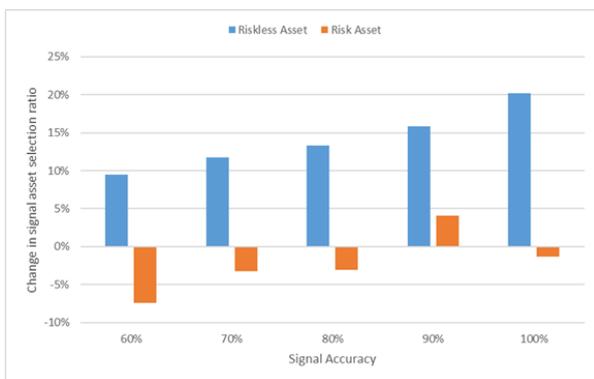

Lastly, we examine cases where signal accuracy varies from one phase to another. We divide the data into two phases: one phase consistently uses a signal of Accuracy 60%, while the other phase employs Accuracy ranging from 100% to 50%, which is added to the state variables.

In Fig. 6, the momentum generated by the past returns of the non-risky asset is applied as a dummy phase for the state variable. The horizontal axis illustrates the difference in Accuracy, whereas the vertical axis indicates the difference between the two phases for the selected proportion of the asset in a state suggesting outperformance. As the horizontal axis shifts rightward and the accuracy in one aspect becomes higher, the proportion of selecting the asset indicated by the signals increases. This outcome is confirmed for both risky and non-risky assets. These results indicate that individuals are able to learn the appropriate aspect to act upon with confidence, even though the difference in signal accuracy is not explicitly given.

Figure 6: Signal accuracy difference and asset selection ratio

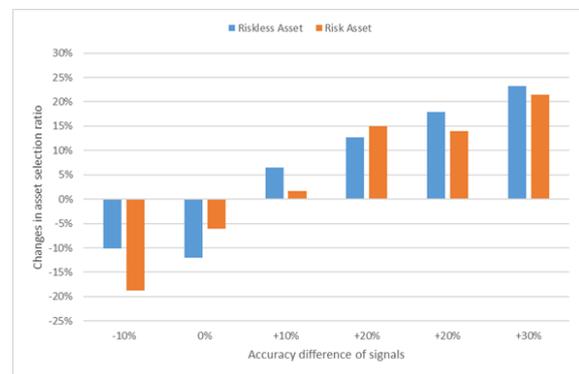

## 6. Summary

This study examines the optimization of dynamic asset allocation using reinforcement learning. The following are the conclusions drawn from this research.

Firstly, it is suggested that incorporating variables related to phase change into the state space of reinforcement learning enhances the out-of-sample performance.

Secondly, it is demonstrated that decision-making alters by incorporating practical constraints into the reinforcement learning conditions and rewards. Further details are as follows.

| 1 | When targets are achieved during a period, risk-averse behavior tends to occur towards the end of the period. |
| 2 | Conversely, if the objective is not accomplished, more aggressive risk-taking is preferred towards the end of the period. |
| 3 | It is learned as 1 and 2 without explicitly giving the volatility of each asset. |
| 4 | Even if the first-stage objective is achieved in the case of a two-stage target, decision-making that favors risk-taking for obtaining additional returns is retained. |

| 5 | Enforcing a drawdown threshold and imposing penalties result in risk-averse behavior toward avoiding surpassing the threshold. |
|---|---|
| 6 | In the case of a two-stage drawdown threshold, risk-taking is taken to recover if it is exceeded in the second stage. |
| 7 | In situations where both target and drawdown are defined, risk-averse decisions are favored towards avoiding exceeding the drawdown threshold during the first half of the period, with risk-taking becoming active in favor of achieving the goal in the second half. |
| 8 | Moreover, performance deteriorates with longer intervals between rebalancing frequencies. |
| 9 | The less accurate the signal, the more conservative decision-making becomes even without its explicit specification. |
| 10 | If signal accuracy varies from one phase to another, it is learned which phase has the highest accuracy, increasing the confidence level of action in the phase with high accuracy, even if it is not explicitly provided. |

## 7. Future Issues

In the future, we would like to analyze the behavior of the Category 2 portfolio management rules defined in Chapter 5 by adding specific practical settings other than rebalancing constraints.

## Notes

The contents and views of this paper belong to the authors and are not the official views of their companies.